# Angular dependent vortex pinning mechanisms in YBCO coated conductors and thin films

L. Civale, B. Maiorov, A. Serquis, J. O. Willis, J. Y. Coulter, H. Wang, Q.X. Jia, P.N. Arendt,

J.L. MacManus-Driscoll*, M.P. Maley, and S.R. Foltyn

Superconductivity Technology Center, Los Alamos National Laboratory, Los Alamos, NM 87545

*On leave from Dept. of Materials Science and Metallurgy,

Univ. of Cambridge, Pembroke St., Cambridge CB2 3QZ, U.K.

We present a comparative study of the angular dependent critical current density in $YBa_2Cu_3O_7$ films deposited on IBAD MgO and on single crystal MgO and $SrTiO_3$ substrates. We identify three angular regimes where pinning is dominated by different types of correlated and uncorrelated defects. We show that those regimes are present in all cases, indicating that the pinning mechanisms are the same, but their extension and characteristics are sample dependent, reflecting the quantitative differences in texture and defect density. In particular, the more defective nature of the films on IBAD turns into an advantage as it results in stronger vortex pinning, demonstrating that the critical current density of the films on single crystals is not an upper limit for the performance of the IBAD coated conductors.

The improvement of the critical current density ($J_c$) of YBa$_2$Cu$_3$O$_7$ (YBCO) films on polycrystalline substrates, known as coated conductors (CC), is a topic of enormous technological importance. Traditionally, the highest YBCO $J_c$s have been obtained for thin films on single crystal substrates (SCS), with values that exceed by orders of magnitude those measured in YBCO single crystals. This difference is due to the much larger density of defects in the films, which act as strong vortex pinning centers, although the exact influence of the different types of defects on $J_c$ is still an open problem. So, the goal for CC has been to obtain $J_c$s as large as for films on SCS. However, although the progress in recent years has been continuous and significant, it was always found that the $J_c$ of CC was lower than what could be obtained for a film of the same thickness deposited on SCS.[1] The inferior performance of the CC was due to the diminished $J_c$ at the low angle grain boundaries of the material[2,3] as compared to the intra-grain $J_c$, a fact that, at least qualitatively, has been well understood since the pioneer work on YBCO films on bi-crystals.[4]

This situation was modified by a recent study showing that CC made by pulsed laser deposition (PLD) on polycrystalline Hastelloy, using an MgO template grown by ion beam assisted deposition (IBAD), have $J_c$s at 75.5 K and self field as large as those of films of the same thickness on SCS.[5] The main reason for the improvement is the better texture of the IBAD template, which leads to an in-plane texture of the YBCO (defined by the full width at half maximum [FWHM] of the X-ray $\phi$ scan peak), better than 3°, low enough to preclude the detrimental effect of the higher-angle grain boundaries on $J_c$. This result implies that the current carrying capability of these CC is limited by intra-grain vortex pinning, with the current flowing approximately uniformly across the film width rather than through a percolative path of weak links. This opens up the possibility to use simple macroscopic transport measurements to compare and contrast the temperature ($T$) and field ($\boldsymbol{H}$) dependence of $J_c$ in these CC and in films on SCS, to investigate whether or not the same defects are controlling pinning in both cases. Particularly valuable information can be obtained from the dependence of $J_c$ on the orientation of $\boldsymbol{H}$, as it allows us to discriminate between uncorrelated or correlated disorder and, in the latter case, to determine the orientation of the extended pinning structures.

In this letter we explore the angular dependence of $J_c$ in YBCO films deposited on IBAD MgO and on single crystal MgO and STO substrates. The YBCO films were grown by PLD as described elsewhere.[5] The films on MgO (both IBAD and SCS) have a buffer layer, whose role on the superconducting properties has been described in Ref. 6. Several parameters of the films are collected in Table I. Transport $J_c$ measurements were performed on bridges of ~ 100 to 250 μm width using a four-probe technique and a 1 μV/cm voltage criterion, with the films immersed in liquid $N_2$ (75.5 K). Angular studies were performed in a 7 T split-coil horizontal magnet, with the sample rotating around a vertical axis and $\mathbf{J} \perp \mathbf{H}$ (maximum Lorentz force configuration). The angle $\Theta$ between $\mathbf{H}$ and the normal to the films (which coincides with the crystallographic c axis) was determined to better than 0.1°.

In Fig. 1 we plot $J_c$ as a function of $\Theta$ at several $H$ for three of the samples. Panels (a), (b) and (c) show films #1 (on crystal STO), #2 (on crystal MgO) and #6 (on IBAD MgO), respectively. For clarity, data for only a few of the fields measured are shown. The main goal of this work is to present a comparative analysis of these curves, identifying the nature of the pinning mechanisms that dominate in different $H$-$\Theta$ regions. We must emphasize that the absolute values of $J_c$ cannot be directly compared in films of different thickness, as it is well known that the $J_c$ of films on both SCS and CC is thickness dependent.[7,8] It is not our purpose to investigate this effect; instead we will focus on the dependences of $J_c$ on $H$ and $\Theta$. The first observation is that, overall, the three sets of curves look similar, suggesting that the same types of pinning structures may be at play. For a given field, $J_c$ exhibits a broad maximum centered at $\mathbf{H}$//ab, and a smaller one centered at $\mathbf{H}$//c. These features are typical of YBCO and have been widely observed.[9-10] At high $H$, an additional sharp peak at the ab planes is always present, superimposed on the broad one (e.g., see Fig. 1(a)). All the other films show the same qualitative features. We now discuss the origin of these three features that characterize $J_c(H,\Theta)$, namely the broad peaks at $\mathbf{H}$//ab and $\mathbf{H}$//c and the sharp peak at $\mathbf{H}$//ab.

A source of angular variation in $J_c$ is the electronic mass anisotropy of YBCO. A very useful approach to describe this effect is the *anisotropic scaling approach* of Blatter *et al.*[11] According to this scheme, if pinning is due to *random defects* (uncorrelated disorder), then $J_c^{rd}$ depends on $H$ and $\Theta$ only through a single variable, $J_c^{rd}(H,\Theta) = J_c^{rd}(\widetilde{H})$, where $\widetilde{H} = H\varepsilon(\Theta)$ is the scaled field, $\varepsilon(\Theta) = \left(\cos^2\Theta + \gamma^{-2}\sin^2\Theta\right)^{1/2}$, and $\gamma \sim 5$−7 is the mass anisotropy. Thus, in those $H$-$\Theta$ regions where only random pinning is present the $J_c(H,\Theta)$ curves should collapse onto a single one if plotted as a function of $\widetilde{H}$.[12] This approach only defines the correct variable; it does not provide the explicit function $J_c^{rd}\left(\widetilde{H}\right)$, which depends on the details of the defects.

A simple test shows that this model gives a good description of our data at least in some $H$-$\Theta$ regions. If we select two pairs of $(H,\Theta)$ that produce the same $\widetilde{H}$, such as (1T, 62°) and (2T, 81°), both giving $\mu_0\widetilde{H} = 0.5T$, we find that indeed both $J_c$s are identical within the resolution, as indicated by the arrows in Fig. 1(a). To determine over which $H$-$\Theta$ regions the scaling approach is valid, we took the $J_c\left(\Theta\right)$ curves for each $H$ and plotted them as $J_c$ vs $\widetilde{H}$. We found that, using $\gamma = 5$ as the only adjustable parameter, all the data collapsed onto a single smooth decreasing curve, except for those points corresponding to the angular regions close to $H$//ab and $H$//c. The portion of the curve containing the collapsed data, which represents the random defect contribution $J_c^{rd}\left(\widetilde{H}\right)$, can be mapped back to the $J_c$ vs. $\Theta$ plot for any $H$. The details of the procedure will be described elsewhere.[13] An example for $\mu_0H$ = 5 T is shown in Fig. 1(a) (solid line). This shows that the smooth variation of $J_c\left(\Theta\right)$ over a wide angular range, with the broad maximum at $H$//ab, can be accounted for by the combined effect of random defects and mass anisotropy. As expected, the fit fails to account for either the c-axis peak or the sharp ab-plane peak, indicating that these arise from correlated pinning structures whose effects cannot be described by the anisotropic scaling. This procedure can be applied to all of the samples, as a tool to determine the boundaries of the $H$-$\Theta$ regions where random pinning dominates.

We now turn to the sharp ab-plane peak. Although less apparent in some cases, this extra peak is always present, as shown for instance in the blow up of the $\mu_0H$ = 7 T data for film #6 in the inset of Fig. 1(c). As seen there, $J_c\left(\Theta\right)$ departs from the scaling-approach fit to the random defect contribution $J_c^{rd}$ (dotted line), over a few degrees. A well-known source of enhanced $J_c$ near the ab planes is the intrinsic pinning associated with the periodic modulation of the superconducting order parameter along the c axis, which arises from the layered structure of YBCO.[14] According to the theory, if $H$ is progressively tilted away from the ab planes, vortices are expected to remain locked in the ab direction until the tilt exceeds a certain small *lock-in angle*, and then to form "staircases", with segments locked in between the Cu-O planes connected by kinks, until a larger *trapping angle* is reached.[11] Here we adopt a phenomenological approach and characterize this peak by a single parameter, the FWHM angle $w_{ab}$, taking the value of the $J_c^{rd}$ fit for **H**//ab as the peak's base. An analysis of the *H-T* dependence of $w_{ab}$ will be presented elsewhere; here we just include the $w_{ab}$ values at $\mu_0 H = 7$ T in Table I. We expect the intrinsic pinning effects to be smeared out over an angular range of at least the out-of-plane mosaic spread of the sample. This is indeed the case, as seen in Table I: $w_{ab} \sim 2°$ for the best-aligned films ($\Delta\omega \le 1.2°$) and increases for the films with poorer out-of-plane texture. Several observations are appropriate at this point. First, the presence of this additional sharp peak confirms the conclusion that the broad maximum at **H**//ab is due to mass anisotropy, and not to intrinsic pinning. Second, in addition to the laminar crystal structure, several types of extended defects parallel to the ab planes such as film-substrate interface or intergrowth layers may contribute to this peak. Third, the absence of any systematic difference between IBAD and SCS strongly suggests that those correlated defects are similar in all our PLD films. Fourth, it would be useful to compare these results with those for films grown by methods that produce laminar structures with a proliferation of extended defects in the ab direction.

The remaining pinning regime is the c-axis peak, which is also due to correlated defects. Natural candidates are twin planes[15,16] as well as edge and screw dislocations;[17,18] the peak probably arising from a combination of them. Their angular range of influence is the angle where the actual $J_c(\Theta)$ merges with the random pinning curve $J_c^{rd}(\Theta)$ obtained above. However, such merging is asymptotic and hard to determine, and in addition it depends on a fit. So, we again adopt a simpler and model-independent approach and parameterize the c-axis peaks by their angular width $w_c$, defined as the angle of the minimum in $J_c(\Theta)$, and by their strength $\Delta J_c$, defined as the difference between $J_c(\Theta = 0)$ and $J_c$ at the minimum, see Fig. 1(b). In Fig. 2(a) we plot $w_c(H)$. The qualitative similarity of the curves is apparent; they all initially increase with *H*, reach a maximum around 1 T, and then decrease following similar trends with *H*. In most cases the agreement is even quantitative, with a maximum $w_c \sim 38°$-42°, and 7 T values $\sim 16°$-20°. Films #1 and #3 have smaller $w_c$. With regards to $\Delta J_c$, the examples shown in Fig. 1 indicate that it is more sample dependent than $w_c$, so in Fig. 2(b) we plot a normalized field dependence $\Delta J_c(H)/J^*$. Again we observe the same trends in all films; $\Delta J_c$ increases up to ~ 1 T and then decreases. We found that above 1 T the field dependences are well described by $\Delta J_c = J^* \exp(-H/H_0)$ with $\mu_0 H_0 \sim 1.6 \pm 0.2$ T, and we used these fits to obtain the normalization values $J^*$ for each sample. The qualitative similarities and quantitative differences in Fig. 2 can be understood if the correlated defects that give rise to the peak are the same in all cases, regardless of the substrate, but their densities vary from sample to sample.

At high fields, $w_c$ and $\Delta J_c$ are expected to decrease with $H$ as indeed is observed in Fig. 2, due to the reduction of the fraction of vortices pinned by the extended defects, and the stiffening of the flux lattice.[11] However, at low $H$, where all the vortices are pinned by the correlated defects, the simplest scenario predicts a trapping angle $\varphi_t \sim \sqrt{2\varepsilon_p/\varepsilon_l}$ and a $\Delta J_c$ independent of $H$, in contradiction with the low-field decrease in Fig. 2. Here $\varepsilon_p$ is the pinning energy per unit length of an extended defect and $\varepsilon_l$ is the vortex line tension. One possibility is that the decrease below 1 T simply indicates that other defects become dominant. Indeed, the maximum in $w_c$ roughly coincides with the crossover between the typical dependence $J_c(H) \propto H^{-\alpha}$, with $\alpha \sim 0.5 - 0.6$,[19,20] and the exponential decay at high $H$. A more complex possibility is that the maximum indicates a matching effect, as has been observed in twin boundaries and columnar defects.[15] In any case, the maximum $w_c$ provides a lower estimate $\varepsilon_p/\varepsilon_l \sim 0.2 - 0.3$, much larger than typical values for twin boundaries in YBCO single crystals at this temperature.[16]

Finally, we compare quantitatively the $J_c(H, \Theta)$ in the various samples. In Fig. 3 we show $J_c(\Theta)/J_c^{sf}$ at $\mu_0 H = 1$ T for all the films studied. The $J_c$s are normalized by $J_c^{sf}$ to account for the thickness dependence when comparing films of different thicknesses.[7,8] However, it is important to note that four of the films, namely #1, #2, #3 and #6, have the same $J_c^{sf}$ within 20%, so for those samples Fig. 3 essentially provides a comparison of the absolute $J_c$s. It is clear that the films on IBAD substrates systematically exhibit higher $J_c$ than those on SCS in almost the entire angular range. Results at higher $H$ show similar trends. This is consistent with the studies of defect densities using X-ray diffraction where we have found that YBCO films on IBAD MgO presents as 5 times higher screw-type threading dislocation density than on crystal MgO. Also, RBS channeling indicates that IBAD films have more point defects, dislocation and misorientations.[21] From here we conclude that, once the detrimental effects of the grain boundaries have been eliminated by improvement of the IBAD MgO in-plane texture, the more defective nature of the films on IBAD turns into an advantage as it results in stronger vortex pinning.

In summary, we identified three angular regimes where pinning is dominated by different types of disorder in YBCO films. We showed that these pinning mechanisms are the same in films on single crystal substrates and on high quality CC on well textured IBAD MgO. Over large field and angular ranges $J_c$ is higher in the CC, demonstrating that the $J_c$ of the films on SCS is not an upper limit for their performance.

We thank Yuan Li for her contribution to sample preparation. This work was performed under the auspices of the U.S. Dept. of Energy.

**Caption Table I**: Parameters of the YBCO films measured. Type of substrate, Thickness (t), FWHM in-plane-texture ($\Delta\phi$), FWHM out-of-plane texture ($\Delta\omega$), critical temperature ($T_c$), critical density current at self field, FWHM of the sharp ab plane peak.

**Caption Fig. 1:** Angular dependence of $J_c$ at several fields, for YBCO films grown on (**a**) single crystal STO; (**b**) single crystal MgO; and (**c**) IBAD MgO. The solid line in (a) is the random pinning contribution to $J_c$ at $\mu_0 H = 5$ T, as obtained from the *anisotropic scaling approach*. The arrows and horizontal line are a test for that scaling, see text. The definitions of the height ($\Delta J_c$) and width ($w_c$) of the c axis peak are sketched in (b). **Inset of (c):** Expanded view for $\mu_0 H = 7$ T near ***H***//ab.

**Caption Fig. 2: (a)** Width of the c-axis peak, for all the films measured. **(b)** Normalize height of the c-axis peak. The solid line is the function $\Delta J_c / J^* = \exp(-H/H_0)$, with $\mu_0 H_0 = 1.6 \pm 0.2$, obtained from the average of the fits of all the samples for $\mu_0 H > 1$ T.

**Caption Fig. 3:** Angular dependence of the normalized critical current density for $\mu_0 H = 1$ T. Films grown on SCS are shown in full symbols, while those grown on IBAD are shown in open symbols.

| Sample number | Substrate | Thickness [$\mu$m] | $\Delta\phi$ [deg] | $\Delta\omega$ [deg] | $T_c$ [K] | $J_c^{sf}$ [MA/cm$^2$] | $w_{ab}$ [deg] |
|---|---|---|---|---|---|---|---|
| 1 | Xtal STO | 1.5 | - | 0.29 | 91.7 | 2.35 | 1.6 |
| 2 | Xtal MgO | 1.1 | 3.2 | 1.4 | 88.5 | 1.91 | 5.4 |
| 3 | IBAD MgO | 1.6 | 3.3 | 1.25 | 88.5 | 2.05 | 6.7 |
| 4 | IBAD MgO | 1.65 | 4.1 | 1.25 | 87.5 | 1.70 | - |
| 5 | IBAD MgO | 1.9 | 4.7 | 2.0 | 87.6 | 1.50 | 6.0 |
| 6 | IBAD MgO | 2.3 | 2.1 | 1.1 | 91.0 | 2.17 | 2.7 |
| 7 | IBAD MgO | 4.3 | 2.7 | 0.91 | 89.0 | 0.97 | 2.1 |
| 8 | IBAD MgO | 5.0 | 2.1 | 1.1 | 91.0 | 1.22 | 2.4 |

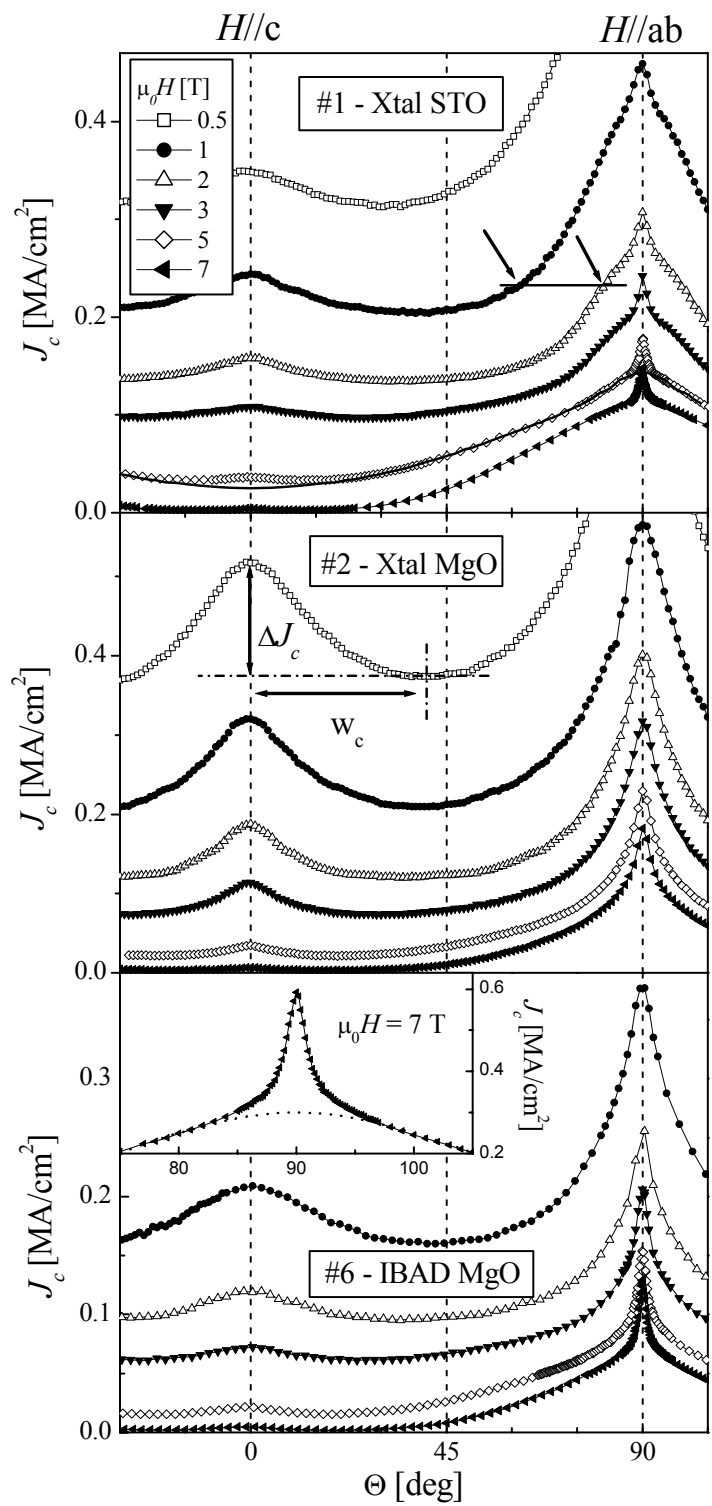

Figure 1 Civale *et al*.

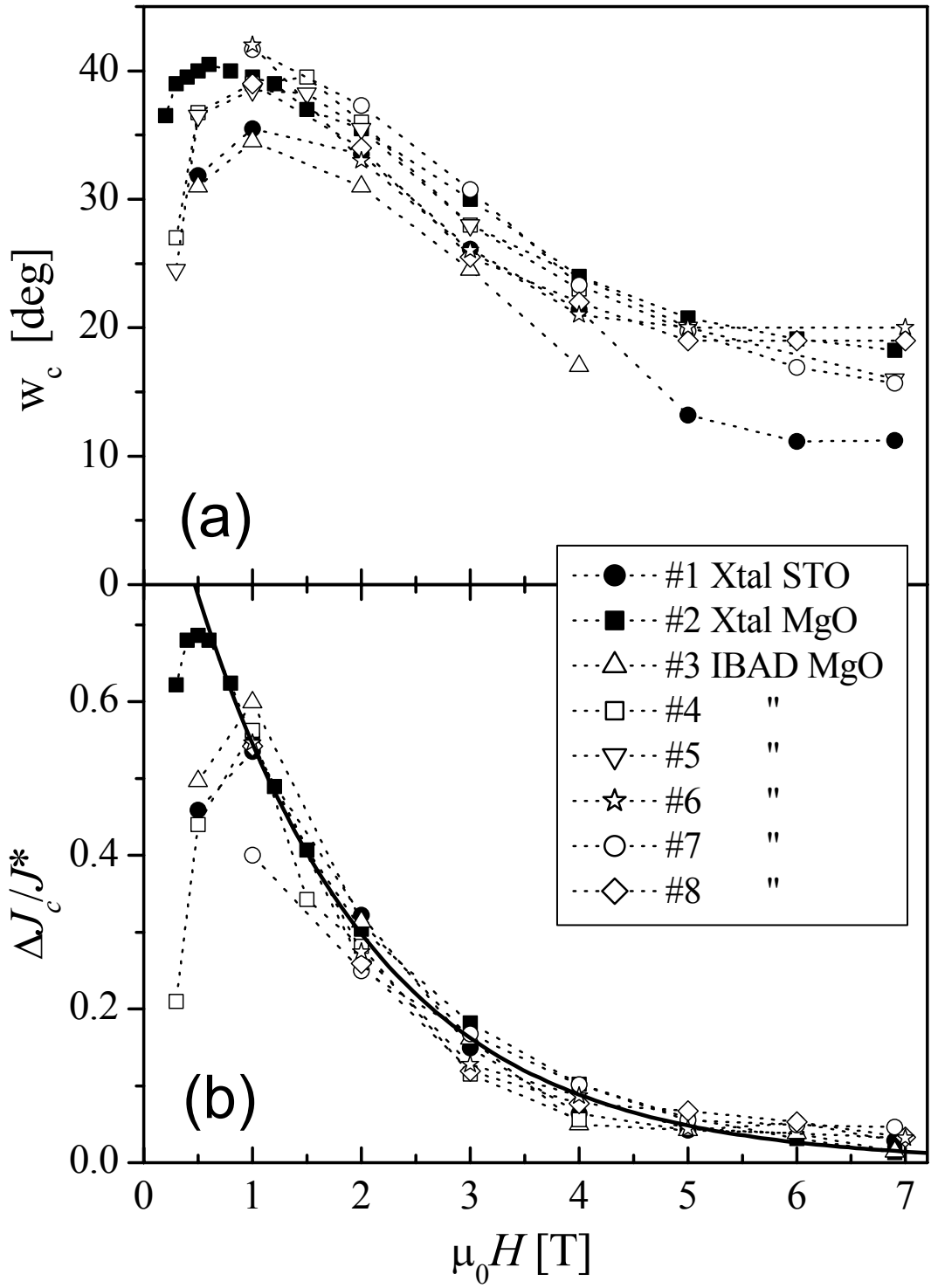

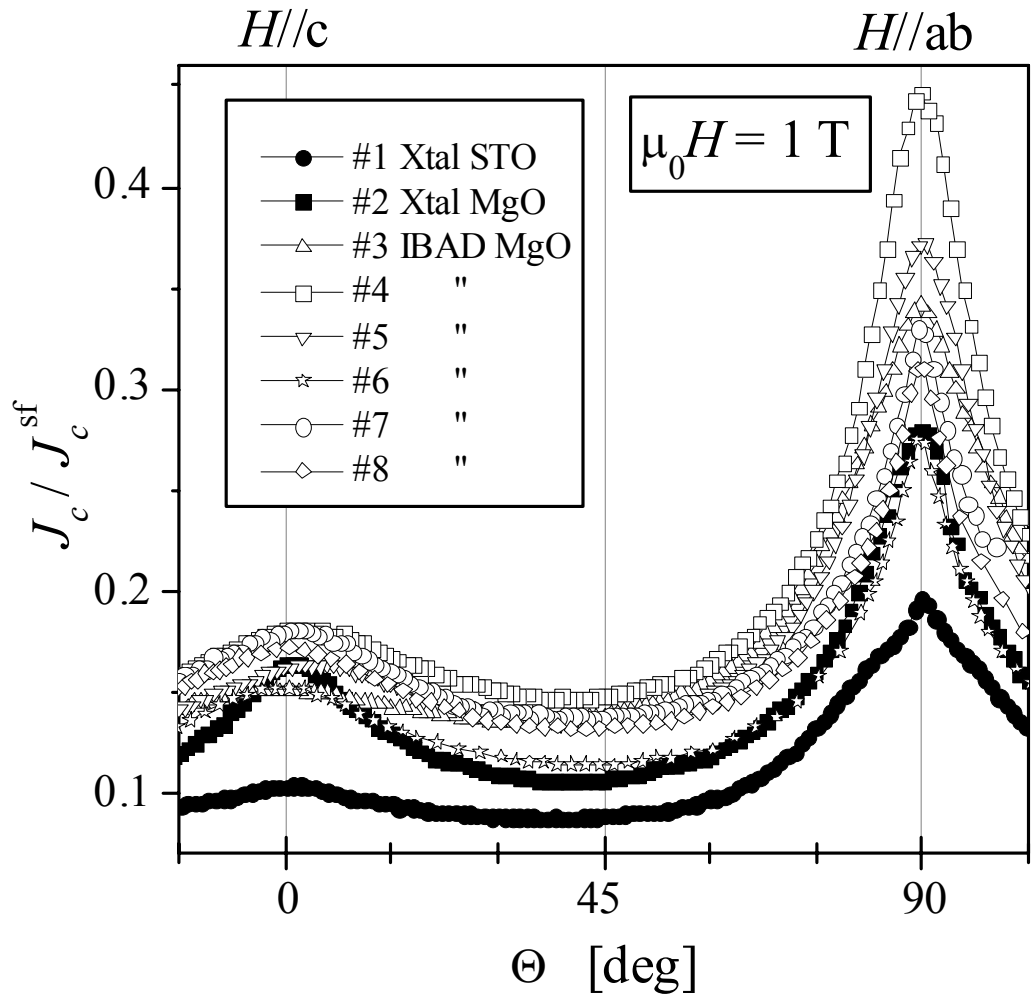